\documentclass[twocolumn,amsmath,nofootinbib]{revtex4} 

\setcitestyle{round}
\usepackage{url}
\usepackage{amssymb}
\usepackage{amsfonts}
\usepackage{amsbsy}
\usepackage{hyperref}
\usepackage{amssymb}
\usepackage{mathrsfs}
\usepackage{amsmath}
\usepackage{mathtools}
\usepackage{algorithm}
\usepackage{algorithmic}
\usepackage{tabularx}
\usepackage{rotating}
\usepackage{enumitem}
\usepackage{mathtools}
\usepackage{courier}

\newcommand{\p}[2]{\frac{\partial #1}{\partial #2}}

\def\ie{{\frenchspacing\it i.e.}}
\def\eg{{\frenchspacing\it e.g.}}
\def\etc{{\frenchspacing\it etc.}}
\def\rms{{\frenchspacing r.m.s.}}

\def\A{\textbf{A}}

\def\D{D}

\def\Eff{\mathcal{F}}

\def\M{\textbf{M}}

\def\U{\textbf{U}}

\def\a{\textbf{a}}
\def\b{\textbf{b}}

\def\dmax{d_{\rm max}}
\def\dmax{d_{\rm max}}
\def\esep{\epsilon_{\rm sep}}
\def\esym{\epsilon_{\rm sym}}
\def\ebr{\epsilon_{\rm br}}
\def\epol{\epsilon_{\rm pol}}
\def\enn{\epsilon_{\rm NN}^{\rm 0}}
\def\epolsep{\epsilon_{\rm pol}^{\rm sep}}
\def\ebfsep{\epsilon_{\rm bf}^{\rm sep}}

\def\frms{f_{\rm rms}}

\def\p{\textbf{p}}

\def\tmax{t_{\rm max}}
\def\u{\textbf{u}}

\DeclareMathAlphabet\mathbfcal{OMS}{cmsy}{b}{n}

\def\x{\textbf{x}}

\def\y{\textbf{y}}

\usepackage{amsmath}

\def\spose#1{\hbox to 0pt{#1\hss}}
\def\simlt{\mathrel{\spose{\lower 3pt\hbox{$\mathchar"218$}}
   \raise 2.0pt\hbox{$\mathchar"13C$}}}
\def\simgt{\mathrel{\spose{\lower 3pt\hbox{$\mathchar"218$}}
     \raise 2.0pt\hbox{$\mathchar"13E$}}}
 \def\simpropto{\mathrel{\spose{\lower 3pt\hbox{$\mathchar"218$}}
     \raise 2.0pt\hbox{$\propto$}}}

\def\beq#1{\begin{equation}\label{#1}}
\def\eeq{\end{equation}}
\def\beqa#1{\begin{eqnarray}\label{#1}}
\def\eeqa{\end{eqnarray}}

\def\fig#1{Figure~\ref{#1}}
\def\Fig#1{Figure~\ref{#1}}

\def\Sec#1{Section~\ref{#1}}

\setcitestyle{square}

\parskip=\medskipamount
\begin{document}
\title{AI Feynman: a Physics-Inspired Method for Symbolic Regression
}
\author{Silviu-Marian Udrescu, Max Tegmark\footnote{Corresponding author. Email: tegmark@mit.edu}}
\address{Dept.~of Physics \& Center for Brains, Minds \& Machines, Massachusetts Institute of Technology, Cambridge, MA 02139; sudrescu@mit.edu}
\address{Theiss Research, La Jolla, CA 92037, USA}
\begin{abstract}
A core challenge for both physics and artificial intelligence (AI) is {\it  symbolic regression}: finding a symbolic expression that matches data from an unknown function.
Although this problem is likely to be NP-hard in principle, functions of practical interest often 
exhibit symmetries, separability, compositionality and other simplifying properties. In this spirit, we develop a recursive multidimensional symbolic regression algorithm that combines neural network fitting with a suite of physics-inspired techniques.
We apply it to 100 equations from the Feynman Lectures on Physics, and it discovers all of them, while previous publicly available software cracks only 71; for a more difficult physics-based test set, we improve the state of the art success rate from  15\% to 90\%.
\end{abstract}
\date{Published in {\it Science Advances}, 6:eaay2631, April 15, 2020}
\vspace{10mm}	

\maketitle

\section{Introduction}
\label{IntroSec}

In 1601, Johannes Kepler got access to the world's best data tables on planetary orbits, and after 4 years and about 40 failed attempts to fit the Mars data to various ovoid shapes, he launched a scientific revolution by discovering that Mars' orbit was an ellipse \cite{koyre2013astronomical}.
This was an example of {\it symbolic regression}: 
discovering a symbolic expression that accurately matches a given data set.
More specifically, we are given a table of numbers, whose rows are of the form $\{x_1,...,x_n,y\}$ where
 $y=f(x_1,...,x_n)$, and our task is to discover the correct symbolic expression for the unknown mystery function $f$, optionally including the complication of noise.

Growing data sets have motivated attempts to automate such regression tasks, with significant success.
For the special case where the unknown function $f$ is a linear combination of known functions of $\{x_1,...,x_n\}$, 
symbolic regression reduces to simply solving a system of linear equations.
{\it Linear} regression (where $f$ is simply an affine function) is ubiquitous in the scientific literature, from finance to psychology.
The case where $f$ is a linear combination of monomials in $\{x_1,...,x_n\}$ corresponds to linear regression with interaction terms, and to polynomial fitting more generally.
There are  countless other examples of popular regression functions that are linear combinations of known functions, ranging from Fourier expansions to wavelet transforms. 
Despite these successes with special cases, the general symbolic regression problem remains unsolved, and it is easy to see why: If we encode functions as strings of symbols, then the number of such strings grows exponentially with string length, so if we simply test all strings by increasing length, it may take longer than the age of our universe until we get to the function we are looking for.

This combinatorial challenge of an exponentially large search space characterizes many famous classes of problems,
from codebreaking and Rubik's cube to the natural selection problem of finding those genetic codes that produce
the most evoutionarily fit organisms.
This has motivated {\it genetic algorithms} \cite{amil2009statistical, pal2017genetic} for targeted searches in exponentially large spaces, which replace the above-mentioned brute-force search by biology-inspired strategies of mutation, selection, inheritance and recombination; crudely speaking, the role of genes is played by useful symbol strings that may form part of the sought-after formula or program. 
 Such algorithms have been successfully applied to areas ranging from design of 
antennas \cite{lohn2002evolutionary,linden2002optimizing}
and vehicles \cite{yu2003application}
to
 wireless routing \cite{panthong20033g},
vehicle routing  \cite{oh2010genetic},
robot navigation \cite{ram1994using}, 
code breaking  \cite{delman2004genetic},
discovering partial differential equations \cite{luextracting},
investment strategy \cite{bauer1994genetic},
marketing \cite{venkatesan2002genetic},
classification \cite{cava2018learning},
Rubik's cube \cite{mcaleer2018solving},
program synthesis \cite{koza1992genetic}
and
metabolic networks \cite{schmidt2011automated}.
  

The symbolic regression problem for mathematical functions (the focus of this paper) has been tackled with a variety of methods
\cite{McRee:2010:SRU:1830761.1830841,Stijven:2011:SWC:2001858.2002059,kong2018a}, 
including sparse regression \cite{mcconaghy2011ffx, arnaldo2015building, brunton2016discovering, quade2018sparse} and
genetic algorithms \cite{searson2010gptips,dubvcakova2011eureqa}.
By far the most successful of these is, as we will see in \Sec{ResultsSec}, the genetic algorithm outlined in 
\cite{schmidt2009distilling} and implemented in the commercial {\it Eureqa} software \cite{dubvcakova2011eureqa}.
  
 The purpose of this paper is to further improve on this state-of-the-art, using physics-inspired strategies enabled by neural networks.  Our most important contribution is using neural networks to discover hidden simplicity such as symmetry or separability in the mystery data, which enables us to recursively break harder problems into simpler ones with fewer variables.
The rest of this paper is organized as follows. In \Sec{MethodsSec},  we present our algorithm and the six strategies that it recursively combines. In \Sec{ResultsSec}, we present a test suite of regression mysteries and use it to test both {\it Eureqa} and our new algorithm, finding major improvements.
In \Sec{ConclusionsSec}, we summarize our conclusions and discuss opportunities for further progress.

\begin{figure}[phbt]
\centerline{\includegraphics[width=88mm]{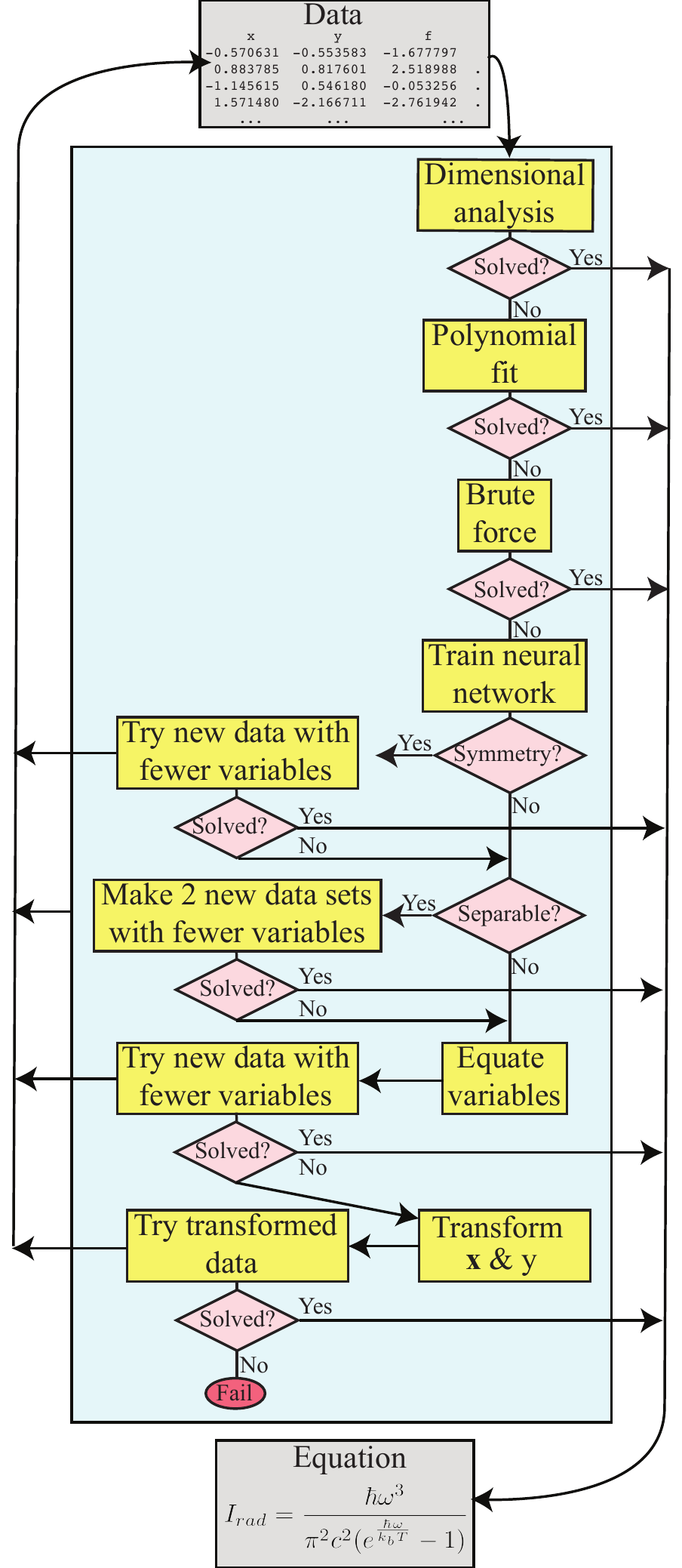}}
\caption{Schematic illustration of our {\it AI Feynman} algorithm. It is iterative as described in the text, with four of the steps capable of generating new mystery data sets that get sent to fresh instantiations of the the algorithm
which may or may not return a solution.
\label{algorithmFig}
}
\end{figure}

\begin{figure}[phbt]
\centerline{\includegraphics[width=80mm]{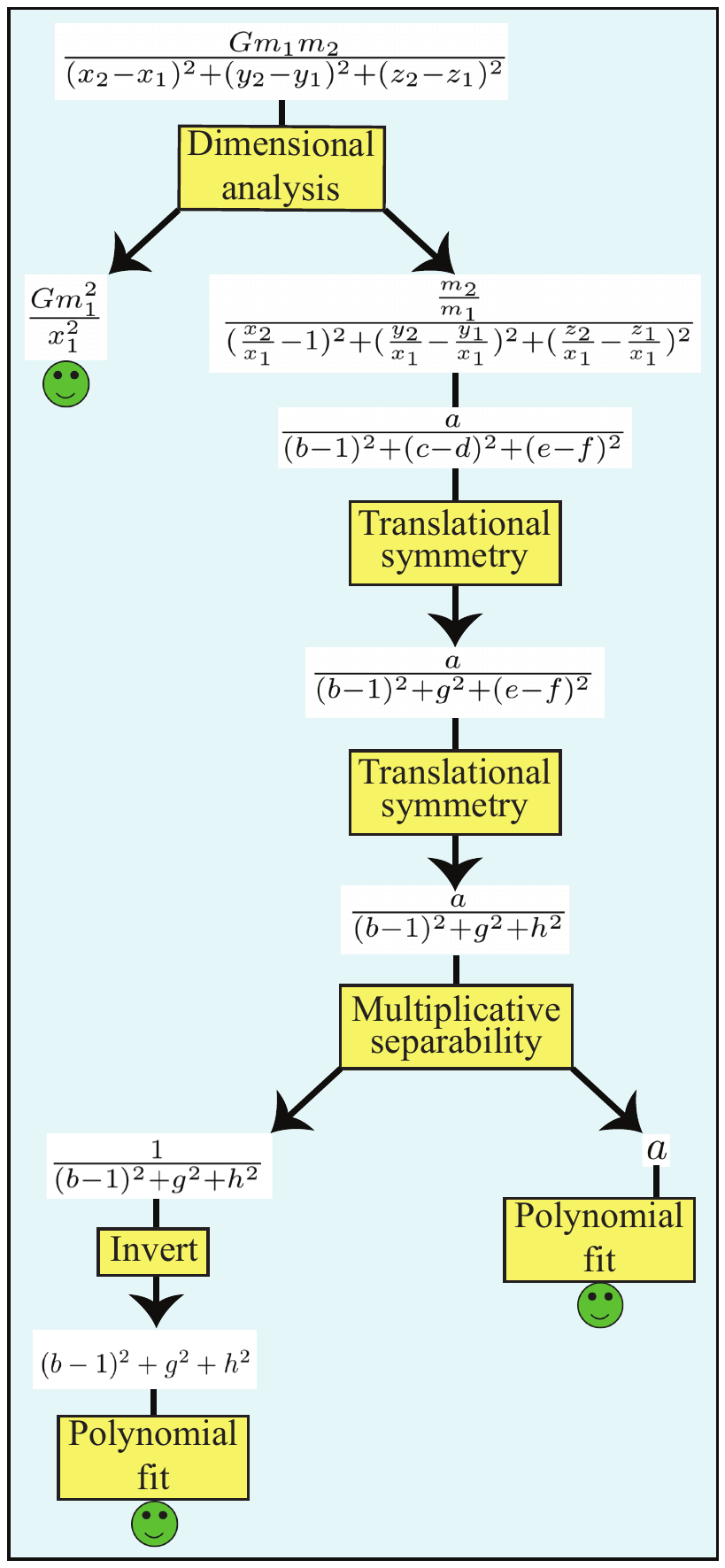}}
\caption{
Example: how our {\it AI Feynman} algorithm discovered mystery Equation~5. Given a mystery table with many examples of the gravitational force $F$ together with the 9 independent variables $G$, $m_1$, $m_2$, $x_1$,..., $z_2$, this table was recursively transformed into simpler ones until the correct equation was found.
First dimensional analysis generated a table of 6 dimensionless independent variables $a=m_2/m_1$, ..., $f=z_1/x_1$
and the dimensionless dependent variable 
$\Eff\equiv F\div G m_1^2/x_1^2$.
Then a neural network was trained to fit this function, which revealed two translational symmetries (each eliminating one variable, by defining $g\equiv c-d$ and $h\equiv e-f$) as well as multiplicative separability,
enabling the factorization $\Eff(a,b,g,h)=G(a)H(b,g,h)$,
thus splitting the problem into two simpler ones.
Both $G$ and $H$ then were solved by polynomial fitting, the latter after applying one of a series of simple transformations (in this case, inversion). For many other mysteries, the final step was instead solved using brute-force symbolic search as 
described in the text.
\label{eq5Fig}
}
\end{figure}

\section{Methods}
\label{MethodsSec}

Generic functions $f(x_1,...,x_n)$ are extremely complicated and near-impossible for symbolic regression to discover. However,  
functions appearing in physics and many other scientific applications often have some of the following simplifying properties that make them easier to discover:
\begin{enumerate}
\itemsep0mm
    \item {\bf Units}: $f$ and the variables upon which it depends have known physical units
    \item {\bf Low-order polynomial}: $f$ (or part thereof) is a polynomial of low degree
    \item {\bf Compositionality:} $f$ is a composition of a small set of elementary functions, each typically taking no more than two arguments
    \item {\bf Smoothness:} $f$ is continuous and perhaps even analytic in its domain
    \item {\bf Symmetry:} $f$ exhibits translational, rotational or scaling symmetry with respect to some of its variables
    \item {\bf Separability:} $f$ can be written as a sum or product of two parts with no variables in common
\end{enumerate}
The question of why these properties are common remains controversial and not fully understood \cite{mhaskar2016learning, cheap}. However, as we will see below, this does not prevent us from discovering and exploiting these properties to facilitate symbolic regression.

Property~(1) enables dimensional analysis, which often transforms the problem into a simpler one with fewer independent variables.
Property~(2) enables polynomial fitting, which quickly solves the problem by solving a system of linear equations to determine the polynomial coefficients.
Property~(3) enables $f$ to be represented as a parse tree with a small number of node types, sometimes enabling $f$ or a sub-expression to be found via a brute-force search.
Property~(4) enables approximating $f$ using a feed forward neural network with a smooth activation function.
Property~(5) can be confirmed using said neural network and enables the problem to be transformed into a simpler one with one independent variable less (or even fewer for $n>2$ rotational symmetry).
Property~(6) can be confirmed using said neural network and enables the independent variables to be partitioned into two disjoint sets, and the 
problem to be transformed into two simpler ones, each involving the variables from one of these sets.

\subsection{Overall Algorithm}

The overall algorithm \footnote{The code is publicly available at 
\url{https://github.com/SJ001/AI-Feynman}.} is schematically illustrated in \fig{algorithmFig}. 
It consists of a series of modules that try to exploit each of the the above-mentioned properties. Like a human scientist, it tries many different strategies (modules) in turn, and if it cannot solve the full problem in one fell swoop, it tries to transform it and divide it into simpler pieces that can be tackled separately, recursively re-launching the full algorithm on each piece. 
\Fig{eq5Fig} illustrates an example of how a particular mystery data set (Newton's law of gravitation with 9 variables) is solved.
Below we describe each of these algorithm modules in turn.

\subsection{Dimensional Analysis}

Our dimensional analysis module exploits the well-known fact that many problems in physics can be simplified by requiring the units of the two sides of an equation to match. This often transforms the problem into a simpler one with a smaller number of variables that are all dimensionless. In the best case scenario, the transformed problem involves solving for a function of zero variables, \ie, a constant. We automate dimensional analysis as follows.

Table \ref{unitTable} shows the physical units of all variables appearing in our 100 mysteries, expressed as products of
the fundamental units (meter, second, kilogram, kelvin, volt) to various integer powers. We thus represent the units of each variable by a vector $\u$ of 5 integers as in the table.
For a mystery of the form $y=f(x_1,...,x_n)$, we define the matrix $\M$ whose $i^{\rm th}$ column
is the $\u$-vector corresponding to the variable $\x_i$, and define the vector $\b$ as the $\u$-vector corresponding to $y$.
We now let the vector $\p$ be a solution to the equation $\M\p = \b$ 
and the columns of the matrix $\U$ form a basis for the null space, so that $\M\U=0$, and define a new mystery
$y'=f'(x'_1,...,x_n')$ where
\beq{prefactorDefEq}
x_i'\equiv \prod_{i=j} ^n {x_j}^{U_{ij}} ,
\quad
y'\equiv {y\over y_*}, 
\quad y_*\equiv \prod_{i=1} ^n {x_i}^{p_i}.
\eeq
By construction, the new variables $x_i'$ and $y'$ are dimensionless, and the number $n'$ of new variables is equal to the dimensionality of the null space.
When $n'>0$, we have the freedom to choose any basis we want for the null space and also to replace $\p$ by a vector of the form 
$\p+\U\a$ 
for any vector $a$; 
we use this freedom to set as many elements as possible in $\p$ and $\U$ equal to zero, \ie, to make the new variables depend on as few old variables as possible.
This choice is useful because it typically results in the resulting powers of the dimensionless variables being integers, making the final expression much easier to find than when the powers are fractions or irrational numbers.

\subsection{Polynomial Fit}

Many functions $f(x_1,...,x_n)$ in physics and other sciences either are low-order polynomials, \eg, the kinetic energy $K=\frac{m}{2}(v_x^2+v_y^2+v_z^2)$, or have parts that are,
 \eg, the denominator of the gravitational force $F=\frac{Gm_1m_2}{(x_1-x_2)^2+(y_1-y_2)^2+(z_1-z_2)^2}$. We therefore include a module that tests if a mystery can be solved by a low-order polynomial.
Our method uses the standard method of solving a system of linear equations to find the best fit polynomial coefficients. It tries fitting the mystery data to polynomials of degree 0, 1, ..., $\dmax=4$ and declares success if the best fitting polynomial gives
r.m.s.~fitting error $\le \varepsilon_p$ (we discuss the setting of this threshold below).

\subsection{Brute Force}

Our brute-force symbolic regression model simply tries all possible symbolic expressions within some class, in order of increasing complexity, terminating either when the maximum fitting error drops below a threshold $\epsilon_p$ or after a maximum runtime $\tmax$ has been exceeded.
Although this module alone could solve all our mysteries in principle, it would in many cases take longer than the age of our universe in practice. Our brute force method is thus typically most helpful once a mystery has been transformed/broken apart into simpler pieces by the modules described below. 

We generate the expressions to try by representing them as strings of symbols, trying first all strings of length 1, then all of length 2, \etc, saving time by only generating those strings that are syntactically correct. 
The symbols used are the independent variables as well a subset of those listed in Table~\ref{opTable}, each representing a constant or a function.
We minimize string length by using reverse Polish notation, so that parentheses become unnecessary. 
For example, $x+y$ can be expressed as 
the string ``\texttt{xy+}'', the number $-2/3$ can be expressed as 
the string ``\texttt{0<<1>>/}" and the relativistic momentum formula $m v/\sqrt{1-v^2/c^2}$ can be expressed as the string ``\texttt{mv*1vv*cc*/-R/}".

\begin{table}[]
\begin{tabular}{|l|l|l|}
\hline
Symbol         & Meaning    &Arguments \\\hline
$+$              & add        	&2 \\
$*$              & multiply    	&2\\
$-$              & subtract    	&2\\
$/$              & divide      	&2\\
\textgreater{} & increment  &1\\
\textless{}    & decrement  &1\\
$\sim$         & negate        &1\\
0              & $0$        &0\\
1              & $1$        &0\\
R              & sqrt       &1 \\
E              & exp       &1\\
P              & $\pi$     &0\\
L              & ln          &1\\
I              & invert     &1\\
C              & cos       &1\\
A              & abs       &1 \\
N              & arcsin   &1\\
T              & arctan    &1\\
S              & sin         &1\\
\hline
\end{tabular}
\caption{\label{opTable}
Functions optionally included in brute force search.
The following three subsets are tried in turn:\\
``+-*/$><\sim$SPLICER'', ``+-*/$>0\sim$" and
\\``+-*/$><\sim$REPLICANTS0".
}
\end{table}

Inspection of Table~\ref{opTable} reveals that many of the symbols are redundant. 
For example, ``\texttt{1}"=``\texttt{0>}'' and ``\texttt{x}$\sim$'' = ``\texttt{0x-}".
$\pi=2\arcsin 1$, so if we drop the symbol ``\texttt{P}", mysteries involving $\pi$ can still get solved with ``\texttt{P}" replaced by ``\texttt{1N1>*}" --- it just takes longer. 

Since there are $s^n$ strings of length $n$ using an alphabet of $s$ symbols, there can be a significant cost both from using too many symbols (increasing $s$) and from using too few symbols (increasing the required $n$, or even making a solution impossible).
As a compromise, our brute force module tries to solve the mystery using three different symbol subsets as explained in the caption of Table~\ref{opTable}.

To exploit the fact that many equations or parts thereof have multiplicative or additive constants, our brute force method comes in two variants that automatically solves for such constants, thus allowing the algorithm to focus on the symbolic expression and not on numerical constants. 

Although the problem of overfitting is most familiar when searching a continuous parameter space, the same phenomenon can occur when searching our discrete space of symbol strings. To mitigate this, we follow the prescription in \cite{aiphysicist} and define
the winning function to be the one with {r.m.s.} fitting error $\epsilon<\epsilon_b$ 
that has the smallest total description length
\beq{DLeq}DL\equiv
\log_2 N
+ \lambda \log_2\left[\max\left(1,{\epsilon\over\epsilon_d}\right)\right],
\eeq
where $\epsilon_d=10^{-15}$ and $N$ is the rank of the string on the list of all strings tried.
The two terms correspond roughly to the number of bits required to store the symbol string and the prediction errors, respectively, if the hyperparameter $\lambda$ is set to equal the number of data points $N_d$. 
We use $\lambda=N_d^{1/2}$ in our experiments below, to prioritize simpler formulas.
If the mystery has been generated using a neural network (see below), we 
set the precision threshold $\epsilon_b$ to ten times the validation error, otherwise we set it to 
$10^{-5}$.

\subsection{Neural-network-based tests \& transformations}

Even after applying the dimensional analysis, many mysteries are still too complex to be solved by the polyfit or brute force modules in a reasonable amount of time. 
However, if the mystery function $f(x_1,...,x_n)$ can be found to have simplifying properties,
it may be possible to transform it into one or more simpler mysteries that can be more easily solved. To search for such properties, we need to be able to evaluate $f$ at points $\{x_1,...,x_n\}$ of our choosing where we typically have no data.
For example, to test if a function $f$ has translational symmetry,
we need to test if 
$f(x_1,x_2)=f(x_1+a,x_2+a)$ for various constants $a$, but 
if a given data point has its two variables separated by $x_2-x_1=1.61803$, we typically have no other examples in our data set with exactly that variable separation. To perform our tests, we thus need an accurate high-dimensional interpolation between our data point.
 
\subsubsection{Neural network training}
In order to obtain such an interpolating function for a given mystery, we train a 
neural network to predict the output given its input. 
We train a feed-forward, fully connected neural network with 6 hidden layers with softplus activation functions, the first 3 having 128 neurons and the last 3 having 64 neurons. 
For each mystery we generated 100,000 data points, using 80\% as the training set and the remainder as the validation set, training for 100 epochs with learning rate 0.005 and batch size 2048. We use the {r.m.s.}-error loss function and the Adam optimizer with a weight decay of $10^{-2}$. The learning rate and momentum schedules were implemented as described in \cite{n.2018superconvergence, smith2018disciplined} using the FastAI package \cite{howard2018fastai}; with a ration of 20 between the maximum and minimum learning rates, and using $10\%$ of the iterations for the last part of the training cycle. For the momentum, the maximum $\beta_1$-value was 0.95 and the minimum 0.85, while $\beta_2=0.99$.

If the neural network were expressive enough to be able to perfectly fit the mystery function, and the training process would never got stuck in a local minimum, then one might naively expect the
{r.m.s.} validation error $\enn$ to scale as 
$\frms\>\epsilon/N_d^{1/2}$ in the limit of ample data, with a constant prefactor depending on the number of function arguments and the function's complexity. Here $\frms$ is the {\rms} of the $f$-values in the dataset,
$N_d$ is the number of data points and $\epsilon$ is the relative {\rms} noise on the independent variable as explored in \Sec{NoiseSec}. For realistic situations, one expects limited expressibility and convergence to keep $\enn$ above some positive floor even as $N_d\to\infty$ and $\epsilon\to 0$.
In practice, we obtained $\enn$-values between $10^{-3}\frms$ and $10^{-5}\frms$ across the range of tested equations.

\subsubsection{Translational symmetry and generalizations}

We test for translational symmetry using the neural network as detailed in Algorithm \ref{tran_sym}.
We first check if the $f(x_1,x_2,x_3,...) =f(x_1+a,x_2+a,x_3...)$ 
to within a precision $\epsilon_{sym}$.
If that is the case, then $f$ depends on $x_1$ and $x_2$ only through their difference, so 
we replace these two input variables by a single new variable $x_1'\equiv x_2-x_1$.
Otherwise, we repeat this test for all pairs of input variables, 
and also test whether any variable pair can be replaced by its sum, product or ratio. 
The ratio case corresponds to scaling symmetry, where two variables can be simultaneously rescaled without changing the answer.  
If any of these simplifying properties is found, the resulting transformed mystery (with one fewer input variables) is iteratively passed into a fresh instantiation of our full {\it AI Feynman} symbolic regression algorithm, as illustrated in \fig{algorithmFig}.
After experimentation, we chose the precision threshold $\epsilon_{sym}$ to be 7 times the neural network validation error, which roughly optimized the training set performance. (If the noise were Gaussian, even a cut at 4 rather than 7 standard deviations would produce negligible false positives.)

\subsubsection{Separability}

We test for separability using the neural network as exemplified in Algorithm~\ref{alg:additive_separability}.
A function is separable if it can be split into two parts with no variables in common. 
We test for both additive and multiplicative separability, corresponding to these two parts being added and multiplied, respectively (the logarithm of a multiplicatively separable function is additively separable).

For example, to test if a function of 2 variables is multiplicatively separable, \ie, of the form 
$f(x_1,x_2)=g(x_1)h(x_2)$ for some univariate functions $g$ and $h$, we first select two constants 
$c_1$ and $c_2$; for numerical robustness, we choose $c_i$ to be the means of all the values of $x_i$ in the mystery data set, $i=1,2$.
We then compute the quantity
\beq{nonSepDefEq}
\Delta_{\rm sep}(x_1,x_2)\equiv \frms^{-1} \left|f(x_1,x_2)-{f(x_1,c_2)f(c_1,x_2)\over f(c_1,c_2)} \right|
\eeq
for each data point. 
This is a measure of non-separability, since it vanishes if $f$ is multiplicatively separable.
The equation is considered separable if the {\rms} average 
$\Delta_{\rm sep}$ over the mystery data set is less than an accuracy threshold $\epsilon_{sep}$, which is chosen to be
$N = 10$ times the neural network validation error \footnote{We also check whether the function is multiplicatively separable up to an additive constant:  $f(x_1,x_2) = a + g(x_1)h(x_2)$, where $a$ is a constant. As a backup, we retain the above-mentioned simpler test for multiplicative separability, which proved more robust when $a=0$}. 

If separability is found, we define the two new univariate mysteries 
$y'\equiv f(x_1,c_2)$
and 
$y''\equiv f(c_1,x_2)/f(c_1,c_2)$. We pass the first one, $y'$, back to a fresh instantiations of our full {\it AI Feynman} symbolic regression algorithm and if it gets solved, we redefine $y''\equiv y/y' c_{num}$, where $c_{num}$ represents any multiplicative numerical constant that appears in $y'$. We then pass $y''$ back to our algorithm and if it gets solved, the final solutions is $y=y'y''/c_{num}$. We test for additive separability analogously, simply replacing $*$ and $/$ by $+$ and $-$ above; also $c_{num}$ will represent an additive numerical constant in this case. If we succeed in solving the two parts,  then the full solution to the original mystery is the sum of the two parts minus the numerical constant. When there are more than two variables $x_i$, we are testing all the possible subsets of variables that can lead to separability, and proceed as above for the newly created two mysteries.

\subsubsection{Setting variables equal}

We also exploit the neural network to explore the effect of setting two input variables equal and attempting to solve the corresponding new mystery $y'$ with one fewer variable. We try this for all variable pairs, and if the resulting new mystery is solved, we try solving 
the mystery $y''\equiv y/y'$ that has the found solution divided out. 

As an example, this technique solves the Gaussian probability distribution mystery I.6.2. After making $\theta$ and $\sigma$ equal, and dividing the initial equation by the result, we are getting rid of the denominator and the remaining part of the equation is an exponential. After taking the logarithm of this (see the below section) the resulting expression can be easily solved by the brute force method.

\subsection{Extra Transformations}

In addition, several transformations are applied to the dependent and independent variables which proved to be useful for solving certain equations. Thus, for each equation, we ran the brute force and polynomial fit on a modified version of the equation in which the dependent variable was transformed by one of the following functions: square root, raise to the power of 2, log, exp, inverse, sin, cos, tan, arcsin, arccos, arctan. This reduces the number of symbols needed by the brute force by one and in certain cases it even allows the polynomial fit to solve the equation, when the brute force would otherwise fail. For example, the formula for the distance between 2 points in the 3D Euclidean space: $\sqrt{(x_1-x_2)^2+(y_1-y_2)^2+(z_1-z_2)^2}$, once raised to the power of 2 becomes just a polynomial which can be easily discovered by the polynomial fit algorithm. The same transformations are also applied to the dependent variables, one at a time. In addition multiplication and division by 2 were added as transformations in this case.

It should be noted that, like most machine-learning methods, the {\it AI Feynman} algorithm has some hyperparameters that can be tuned to optimize performance on the problems at hand. They were all introduced above, but for convenience, they are also summarized in Table~\ref{hyperparameterTable}.

\renewcommand{\arraystretch}{1.3}
\begin{table}[]
\begin{tabular}{|l|p{5.1cm}|c|}
\hline
Symbol         & Meaning     &Setting\\
\hline
$\ebr$              &Tolerance in brute force module    &$10^{-5}$ \\
$\epol$              &Tolerance in polynomial fit module    &$10^{-4}$ \\
$\enn$              &Validation error tolerance for neural network use & $10^{-2}$ \\
$\esep$              & Tolerance for separability        &$10\>\epsilon_{NN}$\\
$\esym$              & Tolerance for symmetry           & $7\>\epsilon_{NN}$\\
$\ebfsep$ &Tolerance in brute force module after separability& $10\>\epsilon_{NN}$\\
$\epolsep$ &Tolerance in polynomial fit module after separability& $10\>\epsilon_{NN}$\\
$\lambda$ &Importance of accuracy relative to complexity& $N_d^{1/2}$\\
\hline
\end{tabular}
\caption{Hyperparameters in our algorithm and the setting we use in this paper.
\label{hyperparameterTable}
}
\end{table}
\renewcommand{\arraystretch}{1}

\section{Results}
\label{ResultsSec}

\subsection{The Feynman Symbolic Regression Database}

To facilitate quantitative testing of our and other symbolic regression algorithms, we created the Feynman Symbolic Regression Database 
(FSReD)
and made it freely available for download\footnote{
The 6.5$\>$GB Feynman Database for Symbolic Regression can be downloaded here:
\url{https://space.mit.edu/home/tegmark/aifeynman.html}}.
For each regression mystery, the database contains the following:
\begin{enumerate}
    \item {\bf Data table:} A table of numbers, whose rows are of the form $\{x_1,x_2,...,y\}$, where $y=f(x_1,x_2,...)$; the challenge is to discover the correct analytic expression for the mystery function $f$.
    \item {\bf Unit table:} A table specifying the physical units of the input and output variables as 6-dimensional vectors of the form seen in Table~\ref{unitTable}.
    \item {\bf Equation:} The analytic expression for the mystery function $f$, for answer-checking.  
\end{enumerate}

To test an analytic regression algorithm using the database, its task is to predict $f$ for each mystery taking the data table (and optionally the unit table) as input. 
Of course, there are typically many symbolically different ways of expressing the same function. For example, if the mystery function $f$ is $(u+v)/(1+uv/c^2)$, then the symbolically different expression $(v+u)/(1+uv/c^2)$ should count as a correct solution. The rule for evaluating an analytic regression method is therefore that a mystery function $f$ is deemed correctly solved by a candidate expression $f'$ if algebraic simplification of the expression $f'-f$ (say, with the \texttt{Simplify} function in {\it Mathematica} or the \texttt{simplify} function in the {\it Python sympy package}) produces the symbol $``0"$.

\begin{table}[]
\begin{tabular}{|l|l|rrrrr|}
\hline
Variables   & Units                   & m  & s  & kg & T  & V  \\
\hline
$a$, $g$          & Acceleration            & 1  & -2 & 0  & 0  & 0    \\
$h$, $\hbar$, $L$, $J_z$ & Angular momentum        & 2  & -1 & 1  & 0  & 0    \\
$A$          & Area                    & 2  & 0  & 0  & 0  & 0    \\
$k_b$         & Boltzmann constant      & 2  & -2 & 1  & -1 & 0    \\
$C$          & Capacitance             & 2  & -2 & 1  & 0  & -2   \\
$q$, $q_1$, $q_2$          & Charge                  & 2  & -2 & 1  & 0  & -1   \\
$j$          & Current density         & 0  & -3 & 1  & 0  & -1   \\
$I$, $I_0$ & Current Intensity       & 2  & -3 & 1  & 0  & -1   \\
$\rho$, $\rho_0$        & Density                 & -3 & 0  & 1  & 0  & 0    \\
$\theta$, $\theta_1$, $\theta_2$, $\sigma$, $n$   & Dimensionless           & 0  & 0  & 0  & 0  & 0    \\
$g\_$, $k_f$, $\gamma$, $\chi$, $\beta$, $\alpha$    & Dimensionless           & 0  & 0  & 0  & 0  & 0   \\
$p_{\gamma}$, $n_0$, $\delta$, $f$, $\mu$       & Dimensionless           & 0  & 0  & 0  & 0  & 0    \\
$n_0$, $\delta$, $f$, $\mu$, $Z_1$, $Z_2$       & Dimensionless           & 0  & 0  & 0  & 0  & 0    \\
$D$          & Diffusion coefficient   & 2  & -1 & 0  & 0  & 0    \\
$\mu_{drift}$  & Drift velocity constant & 0  & -1 & 1  & 0  & 0    \\
$p_d$       & Electric dipole moment  & 3  & -2 & 1  & 0  & -1   \\
$E_f$         & Electric field          & -1 & 0  & 0  & 0  & 1    \\
$\epsilon$    & Electric permitivity    & 1  & -2 & 1  & 0  & -2   \\
$E$, $K$, $U$       & Energy                  & 2  & -2 & 1  & 0    & 0  \\
$E_{den}$     & Energy density          & -1 & -2 & 1  & 0  & 0    \\
$F_E$       & Energy flux             & 0  & -3 & 1  & 0  & 0   \\
$F$, $N_n$          & Force                   & 1  & -2 & 1  & 0  & 0    \\
$\omega$, $\omega_0$      & Frequency               & 0  & -1 & 0  & 0  & 0    \\
$k_G$ &Grav.~coupling ($Gm_1m_2$) &	3 &	-2 & 1 & 0 & 0  \\
$H$ &	Hubble constant &	0 &	-1 & 0 & 0 & 0  \\
$L_{ind}$     & Inductance              & -2 & 4  & -1 & 0  & 2   \\
$n_{rho}$     & Inverse volume          & -3 & 0  & 0  & 0  & 0    \\
$x$, $x_1$, $x_2$, $x_3$    & Length                  & 1  & 0  & 0  & 0  & 0   \\
$y$, $y_1$, $y_2$, $y_3$    & Length                  & 1  & 0  & 0  & 0  & 0   \\
$z$, $z_1$, $z_2$, $r$, $r_1$, $r_2$     & Length                  & 1  & 0  & 0  & 0  & 0   \\
$\lambda$, $d_1$, $d_2$, $d$, $f_f$, $a_f$          & Length                  & 1  & 0  & 0  & 0  & 0   \\
$I_1$, $I_2$, $I_*$, $I_{*_0}$         & Light intensity         & 0  & -3  & 1  & 0  & 0   \\
$B$, $B_x$, $B_y$, $B_z$         & Magnetic field          & -2 & 1  & 0  & 0  & 1    \\
$\mu_m$        & Magnetic moment         & 4  & -3 & 1  & 0  & -1  \\
$M$          & Magnetisation           & 1  & -3 & 1  & 0  & -1   \\
$m$, $m_0$, $m_1$, $m_2$       & Mass                    & 0  & 0  & 1  & 0  & 0    \\
$\mu_e$        & Mobility                & 0  & 1  & -1 & 0  & 0    \\
$p$          & Momentum                & 1  & -1 & 1  & 0  & 0    \\
$G$          & Newton's constant       & 3  & -2 & -1 & 0  & 0    \\
$P_*$        & Polarization            & 0  & -2 & 1  & 0  & -1   \\
$P$        & Power                   & 2  & -3 & 1  & 0  & 0    \\
$p_F$         & Pressure                & -1 & -2 & 1  & 0  & 0    \\
$R$          & Resistance              & -2 & 3  & -1 & 0  & 2    \\
$\mu_S$      & Shear modulus           & -1 & -2 & 1  & 0  & 0    \\
$L_{rad}$     & Spectral radiance       & 0  & -2 & 1  & 0  & 0    \\
$k_{spring}$  & Spring constant         & 0  & -2 & 1  & 0  & 0    \\
$\sigma_{den}$ & Surface Charge density  & 0  & -2 & 1  & 0  & -1   \\
$T$, $T_1$, $T_2$          & Temperature             & 0  & 0  & 0  & 1  & 0    \\
$\kappa$      & Thermal conductivity    & 1  & -3 & 1  & -1 & 0    \\
$t$, $t_1$          & Time                    & 0  & 1  & 0  & 0    & 0  \\
$\tau$        & Torque                  & 2  & -2 & 1  & 0  & 0    \\
$A_{vec}$     & Vector potential        & -1 & 1  & 0  & 0  & 1    \\
$u$, $v$, $v_1$, $c$, $w$  & Velocity                & 1  & -1 & 0  & 0  & 0    \\
$V$, $V_1$, $V_2$          & volume                  & 3  & 0  & 0  & 0  & 0    \\
$\rho_c$, $\rho_{c_0}$     & Volume charge density   & -1 & -2 & 1  & 0  & -1   \\
$V_e$       & Voltage                 & 0  & 0  & 0  & 0  & 1    \\
$k$         & Wave number             & -1 & 0  & 0  & 0  & 0    \\
$Y$          & Young modulus           & -1 & -2 & 1  & 0  & 0    \\
\hline
\end{tabular}
\caption{Unit table used for our automated dimensional analysis.
\label{unitTable}}
\end{table}

In order to sample equations from a broad range of physics areas, the database is generated using  100 equations from the seminal {\it Feynman Lectures on Physics} \cite{feynman1963v1, feynman1963v2, feynman1963v3}, a challenging three-volume course covering classical mechanics, electromagnetism and quantum mechanics as well as a selection of other core physics topics; we prioritized the most complex equations, excluding ones involving derivatives or integrals.
The equations are listed in tables~\ref{EqTableA} and~\ref{EqTableB}, and can be seen to involve between 1 and 9 independent variables as well as the elementary functions $+$, $-$, $*$, $/$, sqrt,
$\exp$, $\log$, $\sin$, $\cos$, $\arcsin$ and $\tanh$. The numbers appearing in these equations are seen to be simple rational numbers as well as $e$ and $\pi$.

We also included in the database a set of 20 more challenging ``bonus" equations, 
extracted from other seminal physics books: Classical Mechanics by Herbert Goldstein, Charles P. Poole, John L. Safko \cite{goldstein2002classical}, Classical electrodynamics by J. Jackson \cite{jackson_classical_1999}, Gravitation and Cosmology: Principles and Applications of the General Theory of Relativity by Steven Weinberg \cite{Weinberg1972-WEIGAC} and Quantum Field Theory and the Standard Model by Matthew D. Schwartz \cite{schwartz2014quantum}. These equations were selected for being both famous and complicated.

The data table provided for each mystery equation contains $10^5$ rows corresponding to randomly generated input variables. These are sampled uniformly between 1 and 5. For certain equations, the range of sampling was slightly adjusted to avoid unphysical result, such as division by zero, or taking the square root of a negative number. The range used for each equation is listed in the Feynman Symbolic Regression Database.

\begin{table*}[]
{\footnotesize
\begin{tabular}{|l|l|r|l|l|l|l|l|}
\hline
Feynman   & Equation & Solution& Methods & Data & Solved  &Solved &Noise\\
eq.   &  & time (s) & used & needed &by Eureqa &w/o da &tolerance\\
\hline                            
I.6.20a       & $f = e^{-\theta^2/2}/\sqrt{2\pi}$ & 16 & bf  & $10$ & no & yes & $10^{-2}$ \\
I.6.20        & $f = e^{-\frac{\theta^2}{2\sigma^2}}/\sqrt{2\pi\sigma^2}$ & 2992 & ev, bf-log & $10^2$ & no & yes  & $10^{-4}$\\
I.6.20b       & $f = e^{-\frac{(\theta-\theta_1)^2}{2\sigma^2}}/\sqrt{2\pi\sigma^2}$ & 4792 &  sym--, ev, bf-log & $10^3$& no & yes & $10^{-4}$ \\
I.8.14       & $d = \sqrt{(x_2-x_1)^2+(y_2-y_1)^2}$ & 544 & da, pf-squared  & $10^2$ & no & yes & $10^{-4}$  \\
I.9.18       & $F = \frac{Gm_1m_2}{(x_2-x_1)^2+(y_2-y_1)^2+(z_2-z_1)^2}$  & 5975 & da, sym--, sym--, sep$*$, pf-inv  & $10^6$ & no & yes  & $10^{-5}$         \\
I.10.7       & $m = \frac{m_0}{\sqrt{1-\frac{v^2}{c^2}}}$   & 14 & da, bf     & $10$  & no & yes  & $10^{-4}$    \\
I.11.19      & $A = x_1y_1+x_2y_2+x_3y_3$ & 184 & da, pf   & $10^2$     & yes     & yes  & $10^{-3}$    \\
I.12.1       & $F = \mu N_n$ & 12 & da, bf  & $10$          & yes & yes  & $10^{-3}$ \\
I.12.2       & $F = \frac{q_1q_2}{4\pi\epsilon r^2}$   & 17 & da, bf  & $10$  & yes & yes & $10^{-2}$\\
I.12.4       & $E_f = \frac{q_1}{4\pi\epsilon r^2}$  & 12 & da     & $10$   & yes & yes & $10^{-2}$  \\
I.12.5       & $F = q_2 E_f$ & 8 & da & $10$            & yes & yes  & $10^{-2}$  \\
I.12.11      & $F = q(E_f+B v \sin\theta)$  & 19 & da, bf & $10$   & yes  & yes  & $10^{-3}$  \\
I.13.4      & $K = \frac{1}{2}m(v^2+u^2+w^2)$  & 22 & da, bf   & $10$   & yes & yes   & $10^{-4}$   \\
I.13.12      & $U = Gm_1m_2(\frac{1}{r_2}-\frac{1}{r_1})$ & 20 & da, bf & $10$ & yes  & yes  & $10^{-4}$ \\
I.14.3       & $U = mgz$ & 12 & da    & $10$        & yes     & yes  & $10^{-2}$ \\
I.14.4       & $U = \frac{k_{spring}x^2}{2}$  & 9 & da   & $10$   & yes    & yes  & $10^{-2}$ \\
I.15.3x      & $x_1 = \frac{x-ut}{\sqrt{1-u^2/c^2}}$ & 22 & da, bf  & $10$    & no & no   & $10^{-3}$ \\
I.15.3t      & $t_1 = \frac{t-ux/c^2}{\sqrt{1-u^2/c^2}}$ & 20 & da, bf  & $10^2$ & no & no & $10^{-4}$  \\
I.15.10       & $p = \frac{m_0v}{\sqrt{1-v^2/c^2}}$ & 13 & da, bf & $10$ & no & yes & $10^{-4}$ \\
I.16.6       & $v_1 = \frac{u+v}{1+uv/c^2}$ & 18 & da, bf   & $10$  & no & yes & $10^{-3}$  \\
I.18.4       & $r = \frac{m_1r_1+m_2r_2}{m_1+m_2}$ & 17 & da, bf  & $10$ & yes & yes & $10^{-2}$ \\
I.18.12      & $\tau = rF\sin\theta$  & 15 & da, bf  & $10$ & yes & yes & $10^{-3}$ \\
I.18.16      & $L = mrv \sin\theta$  & 17 & da, bf  & $10$ & yes & yes & $10^{-3}$ \\
I.24.6 & $E = \frac{1}{4} m (\omega^2+\omega_0^2) x^2$      & 22 & da, bf   & $10$   & yes   & yes & $10^{-4}$\\
I.25.13      & $V_e = \frac{q}{C}$ & 10 & da     & $10$    & yes & yes & $10^{-2}$ \\
I.26.2       & $\theta_1 = \arcsin(n  \sin\theta_2)$ & 530 & da, bf-sin  & $10^2$  & yes & yes & $10^{-2}$ \\
I.27.6       & $f_f$    $ = \frac{1}{\frac{1}{d_1}+\frac{n}{d_2}}$  & 14 & da, bf & $10$ & yes & yes & $10^{-2}$ \\
I.29.4       & $k = \frac{\omega}{c}$ & 8 & da  & $10$  & yes & yes & $10^{-2}$ \\
I.29.16      & $x = \sqrt{x_1^2+x_2^2-2x_1x_2\cos(\theta_1-\theta_2)}$ & 2135 & da, sym--, bf-squared & $10^3$   & no  & no & $10^{-4}$  \\
I.30.3 & $I_* = I_{*_0}\frac{\sin^2(n\theta/2)}{\sin^2(\theta/2)}$ & 118 & da, bf   & $10^2$ & yes  & yes & $10^{-3}$ \\
I.30.5       & $\theta = \arcsin(\frac{\lambda}{nd})$  & 529 & da, bf-sin  & $10^2$ & yes & yes & $10^{-3}$  \\
I.32.5       & $P = \frac{q^2a^2}{6\pi\epsilon c^3}$       & 13 & da & $10$  & yes   & yes & $10^{-2}$ \\
I.32.17 & $P = (\frac{1}{2}\epsilon c E_f^2)(8\pi r^2/3) (\omega^4/(\omega^2-\omega_0^2)^2)$      & 698 & da, bf-sqrt     & $10$    & no  & yes  & $10^{-4}$  \\
I.34.8       & $\omega = \frac{qvB}{p}$   & 13 & da  & $10$  & yes  & yes & $10^{-2}$ \\
I.34.10       & $\omega = \frac{\omega_0}{1-v/c}$ & 13 & da, bf  & $10$  & no  & yes  & $10^{-3}$ \\
I.34.14      & $\omega = \frac{1+v/c}{\sqrt{1-v^2/c^2}}\omega_0$  & 14 & da, bf  & $10$ & no & yes & $10^{-3}$  \\
I.34.27      & $E = \hbar\omega$  & 8 & da  & $10$ & yes  & yes & $10^{-2}$  \\
I.37.4       & $I_* = I_1+I_2+2\sqrt{I_1I_2}\cos\delta$ & 7032 & da, bf  & $10^2$ & yes & no & $10^{-3}$ \\
I.38.12      & $r = \frac{4\pi\epsilon\hbar^2}{mq^2}$   & 13 & da  & $10$  & yes   & yes & $10^{-2}$ \\
I.39.10       & $E = \frac{3}{2}p_F V$     & 8 & da   & $10$  & yes & yes & $10^{-2}$ \\
I.39.11      & $E = \frac{1}{\gamma-1}p_F V$  & 13 & da, bf   & $10$  & yes   & yes  & $10^{-3}$ \\
I.39.22      & $P_F = \frac{n k_b T}{V}$       & 16 & da, bf   & $10$  & yes & yes & $10^{-4}$ \\
I.40.1       & $n = n_0e^{-\frac{mgx}{k_bT}}$    & 20 & da, bf    & $10$   & no  & yes   & $10^{-2}$ \\
I.41.16      & $L_{rad} = \frac{\hbar\omega^3}{\pi^2c^2(e^{\frac{\hbar\omega}{k_bT}}-1)}$ & 22 & da, bf & $10$  & no  & no & $10^{-5}$ \\
I.43.16      & $v = \frac{\mu_{drift}q V_e}{d}$   & 14 & da   & $10$   & yes & yes & $10^{-2}$ \\
I.43.31      & $D = \mu_e k_bT$                            & 11 & da   & $10$   & yes  & yes & $10^{-2}$ \\
I.43.43      & $\kappa = \frac{1}{\gamma-1}\frac{k_bv}{A}$  & 16 & da, bf & $10$  & yes & yes & $10^{-3}$  \\
I.44.4       & $E = n k_b T \ln(\frac{V_2}{V_1})$      & 18 & da, bf  & $10$ & yes   & yes & $10^{-3}$ \\
I.47.23      & $c = \sqrt{\frac{\gamma pr}{\rho}}$   & 14& da, bf  & $10$    & yes   & yes & $10^{-2}$ \\
I.48.20       & $E = \frac{m c^2}{\sqrt{1-v^2/c^2}}$ &  108& da, bf  & $10^2$   & no  & no & $10^{-5}$ \\
I.50.26 & $x = x_1[\cos(\omega t)+\alpha\> cos(\omega t)^2]$      & 29 & da bf       & $10$  & yes  & yes &  $10^{-2}$    \\
\hline
\end{tabular}
\caption{Tested Feynman Equations, part 1. Abbreviations in the ``Methods used" column: ``da" = dimensional analysis, ``bf" = brute force, ``pf" = polyfit, ``ev" = set 2 variables equal, ``sym" = symmetry,  ``sep" = separability. Suffixes denote the type of symmetry or separability (``sym$-$" =translationa symmetry, "sep*"=multiplicative separability, \protect\etc) or the preprocessing before brute force (\protect\eg, ``bf-inverse" means inverting the mystery function before bf). 
\label{EqTableA}
}
}
\end{table*}

\begin{table*}[]
\begin{tabular}{|l|l|r|l|r|l|l|l|}
\hline
Feynman   & Equation & Solution& Methods & Data & Solved  &Solved &Noise\\
eq.   &  & time (s) & used & needed &by Eureqa &w/o DA &tolerance\\
\hline       
II.2.42   & P     $ = \frac{\kappa(T_2-T_1)A}{d}$  & 54 & da, bf & $10$ & yes  & yes & $10^{-3}$ \\
II.3.24   & $F_E = \frac{P}{4\pi r^2}$         & 8 & da  & $10$     & yes & yes & $10^{-2}$  \\
II.4.23   & $V_e = \frac{q}{4\pi\epsilon r}$           & 10 & da  & $10$    & yes      & yes & $10^{-2}$ \\
II.6.11 & $V_e =\frac{1}{4\pi\epsilon}\frac{p_d\cos \theta}{r^2}$      & 18 & da, bf   &  $10$  & yes  & yes & $10^{-3}$   \\
II.6.15a & $E_f = \frac{3}{4\pi\epsilon}\frac{p_d z}{r^5} \sqrt{x^2+y^2}$      & 2801 & da, sm, bf  & $10^4$   & no   & yes & $10^{-3}$ \\
II.6.15b & $E_f = \frac{3}{4\pi\epsilon}\frac{p_d}{r^3} \cos\theta\sin\theta$      & 23 & da, bf   & $10$   & yes   & yes & $10^{-2}$  \\
II.8.7    & $E = \frac{3}{5}\frac{q^2}{4\pi\epsilon d}$    & 10 & da   & $10$     & yes   & yes & $10^{-2}$ \\
II.8.31   & $E_{den} = \frac{\epsilon E_f^2}{2}$                     & 8 & da    & $10$   & yes   & yes & $10^{-2}$ \\
II.10.9   & $E_f = \frac{\sigma_{den}}{\epsilon}\frac{1}{1+\chi}$      & 13 & da, bf   & $10$   & yes   & yes & $10^{-2}$ \\
II.11.3 & $x = \frac{q E_f}{m(\omega_0^2-\omega^2)}$      & 25 & da, bf   & $10$   & yes  & yes & $10^{-3}$       \\
II.11.17 & $n = n_0(1+ \frac{p_d E_f \cos\theta}{k_b T})$      & 28 & da, bf   & $10$   & yes   & yes   & $10^{-2}$     \\
II.11.20  & $P_* = \frac{n_\rho p_d^2 E_f}{3 k_b T}$      & 18 & da, bf   & $10$    & yes   & yes & $10^{-3}$ \\
II.11.27 & $P_* = \frac{n\alpha}{1-n\alpha/3}\epsilon E_f$      & 337 & da  bf-inverse  & $10^2$  & no  & yes & $10^{-3}$      \\
II.11.28  & $\theta = 1+\frac{n\alpha}{1-(n\alpha/3)}$    & 1708& da, sym*, bf & $10^2$  & no   & yes  & $10^{-4}$  \\   
II.13.17  & $B = \frac{1}{4 \pi \epsilon c^2}\frac{2I}{r}$ & 13 & da   & $10$   & yes & yes & $10^{-2}$ \\
II.13.23  & $\rho_c = \frac{\rho_{c_0}}{\sqrt{1-v^2/c^2}}$           & 13 & da, bf  & $10^2$ & no   & yes & $10^{-4}$ \\
II.13.34  & $j = \frac{\rho_{c_0}v}{\sqrt{1-v^2/c^2}}$     & 14 & da, bf    & $10$  & no   & yes & $10^{-4}$ \\
II.15.4   & $E = -\mu_M B \cos\theta$               & 14 & da, bf   & $10$  & yes   & yes & $10^{-3}$ \\
II.15.5   & $E = -p_d E_f\cos\theta$  & 14 & da, bf & $10$ & yes  & yes & $10^{-3}$\\
II.21.32  & $V_e = \frac{q}{4\pi\epsilon r(1-v/c)}$   & 21 & da, bf  & $10$ & yes  & yes  & $10^{-3}$   \\
II.24.17 & $k = \sqrt{\frac{\omega^2}{c^2}-\frac{\pi^2}{d^2}}$      & 62 & da   bf   & $10$    & no  & yes  & $10^{-5}$  \\
II.27.16  & $F_E = \epsilon c E_f^2$        & 13 & da   & $10$  & yes  & yes  & $10^{-2}$ \\
II.27.18  & $E_{den} = \epsilon E_f^2$             & 9 & da  & $10$ & yes & yes   & $10^{-2}$ \\
II.34.2a  & $I = \frac{qv}{2\pi r}$             & 11 & da   & $10$  & yes & yes & $10^{-2}$ \\
II.34.2   & $\mu_M = \frac{q v r}{2}$                       & 11 & da  & $10$ & yes & yes & $10^{-2}$ \\
II.34.11  & $\omega = \frac{g_{\_} q B}{2m}$          & 16 & da, bf & $10$ & yes   & yes & $10^{-4}$ \\
II.34.29a & $\mu_M = \frac{q h}{4\pi m}$      & 12 & da & $10$    & yes  & yes   & $10^{-2}$ \\
II.34.29b & $E = \frac{g_{\_} \mu_M B J_z}{\hbar}$ & 18 & da, bf   & $10$ & yes & yes & $10^{-4}$ \\
II.35.18 & $n = \frac{n_0}{\exp(\mu_m B/(k_b T))+\exp(-\mu_m B/(k_b T))}$      & 30 & da, bf   & $10$   & no   & yes   & $10^{-2}$\\
II.35.21  & $M = n_\rho \mu_M \tanh(\frac{\mu_M B}{k_b T})$     & 1597 & da, halve-input, bf & $10$  & yes  & no & $10^{-4}$ \\
II.36.38 & $f = \frac{\mu_m B}{k_b T}+\frac{\mu_m\alpha M}{\epsilon c^2 k_b T}$      & 77 & da bf  & $10$  & yes  & yes  &  $10^{-2}$ \\
II.37.1   & $E = \mu_M(1+\chi)B$    & 15 & da, bf     & $10$ & yes & yes & $10^{-3}$  \\
II.38.3   & $F = \frac{Y A x}{d}$            & 47 & da, bf   & $10$ & yes & yes  & $10^{-3}$ \\
II.38.14  & $\mu_S = \frac{Y}{2(1+\sigma)}$           & 13 & da, bf & $10$   & yes  & yes & $10^{-3}$  \\
III.4.32  & $n = \frac{1}{e^{\frac{\hbar\omega}{k_bT}}-1}$    & 20 & da, bf   & $10$  & no   & yes  & $10^{-3}$  \\
III.4.33  & $E = \frac{\hbar\omega}{e^{\frac{\hbar\omega}{k_b T}}-1}$  & 19 & da, bf    & $10$   & no  & yes & $10^{-3}$    \\
III.7.38  & $\omega = \frac{2 \mu_M B}{\hbar}$  & 13 & da   & $10$   & yes  & yes & $10^{-2}$   \\
III.8.54  & $p_{\gamma}$    $ = \sin(\frac{E t}{\hbar})^2$    & 39 & da, bf  & $10$  & no & yes  & $10^{-3}$    \\
III.9.52  & $p_{\gamma}$    $ = \frac{p_d E_f t}{\hbar} \frac{    \sin((\omega-\omega_0)t/2)^2}{((\omega-\omega_0)t/2)^2}$ & 3162 & da, sym--, sm, bf & $10^3$ & no & yes & $10^{-3}$  \\
III.10.19 & $E = \mu_M\sqrt{B_x^2+B_y^2+B_z^2}$  & 410 & da, bf-squared  & $10^2$  & yes  & yes  & $10^{-4}$ \\
III.12.43 & $L = n\hbar$ & 11 & da, bf & $10$  & yes  & yes & $10^{-3}$   \\
III.13.18 & $v = \frac{2 E d^2 k}{\hbar}$          & 16 & da, bf   & $10$ & yes   & yes   & $10^{-4}$  \\
III.14.14 & $I = I_0 (e^{\frac{q V_e}{k_b T}}-1)$  & 18 & da, bf  & $10$  & no & yes   & $10^{-3}$  \\
III.15.12 & $E = 2U(1-\cos(kd))$    & 14 & da, bf & $10$ & yes  & yes  & $10^{-4}$  \\
III.15.14 & $m = \frac{\hbar^2}{2E d^2}$     & 10 & da         & $10$    & yes    & yes  & $10^{-2}$   \\
III.15.27 & $k = \frac{2\pi\alpha}{nd}$              & 14 & da, bf   & $10$   & yes   & yes  & $10^{-3}$ \\
III.17.37 & $f = \beta(1+\alpha \cos\theta)$    & 27 & bf   & $10$   & yes   & yes  & $10^{-3}$  \\
III.19.51 & $E = \frac{-mq^4}{2(4\pi\epsilon)^2\hbar^2}\frac{1}{n^2}$     & 18 & da, bf & $10$  & yes  & yes   & $10^{-5}$    \\
III.21.20 & $j = \frac{-\rho_{c_0} q A_{vec}}{m}$      & 13 & da   & $10$   & yes   & yes  &  $10^{-2}$    \\

\hline
\end{tabular}
\caption{Tested Feynman Equations, part 2 (same notation as in Table \ref{EqTableA})
\label{EqTableB}
}
\end{table*}

\renewcommand{\arraystretch}{1.6}
\begin{table*}[]
\begin{tabular}{|l|l|l|l|l|}
\hline
Source & Equation & Solved & Solved by &  Methods used  \\
        &          &        &  Eureqa  &  \\
\hline 
Rutherford Scattering & $A=\left(\frac{Z_1Z_2\alpha\hbar c}{4E\sin^2(\frac{\theta}{2})}\right)^2$  & yes & no & da, bf-sqrt   \\
Friedman Equation & $H=\sqrt{\frac{8\pi G}{3}\rho-\frac{k_f c^2}{a_f^2}}$ & yes & no & da, bf-squared \\
Compton Scattering & $U = \frac{E}{1+\frac{E}{mc^2}(1-\cos\theta)}$  & yes & no & da, bf \\
Radiated gravitational wave power & $P = -\frac{32}{5}\frac{G^4}{c^5}\frac{(m_1m_2)^2(m_1+m_2)}{r^5}$  & no & no & - \\
Relativistic aberration & $\theta_1 = \arccos\left(\frac{\cos\theta_2-\frac{v}{c}}{1-\frac{v}{c}\cos\theta_2}\right)$  & yes & no & da, bf-cos \\
N-slit diffraction & $I=I_0\left[\frac{\sin(\alpha/2)}{\alpha/2}\frac{\sin(N\delta/2)}{\sin(\delta/2)}\right]^2$  & yes & no & da, sm, bf \\
Goldstein 3.16& $v = \sqrt{\frac{2}{m}(E-U-\frac{L^2}{2mr^2})}$  & yes & no & da, bf-squared \\
Goldstein 3.55& $k = \frac{mk_G}{L^2}(1+\sqrt{1+\frac{2EL^2}{mk_G^2}}\cos(\theta_1-\theta_2))$  & yes & no & da, sym-, bf  \\
Goldstein 3.64 (ellipse)& $r=\frac{d(1-\alpha^2)}{1+\alpha\cos(\theta_1-\theta_2)}$  & yes & no & da, sym-, bf \\
Goldstein 3.74 (Kepler)& $t=\frac{2\pi d^{3/2}}{\sqrt{G(m_1+m_2)}}$  & yes & no & da, bf \\
Goldstein 3.99 & $\alpha=\sqrt{1+\frac{2\epsilon^2 EL^2}{m(Z_1Z_2q^2)^2}}$  & yes & no & da, sym*, bf  \\
Goldstein 8.56& $E=\sqrt{(p-qA_{vec})^2c^2+m^2c^4}+qV_e$& yes & no & da, sep$+$, bf-squared \\
Goldstein 12.80& $E=\frac{1}{2m}[p^2+m^2\omega^2x^2(1+\alpha\frac{x}{y})]$ & yes & yes & da, bf \\
Jackson 2.11& $F=\frac{q}{4\pi\epsilon y^2}\left[4\pi\epsilon V_e d-\frac{qdy^3}{(y^2-d^2)^2}\right]$ & no & no & - \\
Jackson 3.45& $V_e=\frac{q}{(r^2+d^2-2dr\cos\alpha)^{\frac{1}{2}}}$ & yes & no & da, bf-inv \\
Jackson 4.60& $V_e=E_f\cos\theta\left(\frac{\alpha-1}{\alpha+2}\frac{d^3}{r^2}-r\right)$ & yes & no & da, sep$*$, bf \\
Jackson 11.38 (Doppler)& $\omega_0=\frac{\sqrt{1-\frac{v^2}{c^2}}}{1+\frac{v}{c}\cos\theta}\>\omega$ & yes & no & da, cos-input, bf \\
Weinberg 15.2.1& $\rho=\frac{3}{8\pi G}\left(\frac{c^2k_f}{a_f^2}+H^2\right)$ & yes & yes & da, bf \\
Weinberg 15.2.2& $p_f=-\frac{1}{8\pi G}\left[\frac{c^4k_f}{a_f^2}+c^2H^2(1-2\alpha)\right]$ & yes & yes & da, bf  \\
Schwarz 13.132 (Klein-Nishina) & $A=\frac{\pi\alpha^2\hbar^2}{m^2c^2}(\frac{\omega_0}{\omega})^2\left[\frac{\omega_0}{\omega}+\frac{\omega}{\omega_0}-\sin^2\theta\right]$ & yes & no & da, sym/, sep*, sin-input, bf \\
\hline
\end{tabular}
\caption{Tested bonus equations. Goldstein 8.56 is for the special case where the vectors $\p$ and $\A$ are parallel.
\label{EqTableC}
}
\end{table*}
\renewcommand{\arraystretch}{1.}

\subsection{Method comparison}

We reviewed the symbolic regression literature for publicly available software against which our method could be compared. To the best of our knowledge, the best competitor by far is the commercial {\it Eureqa} software sold by Nutonian, Inc.\footnote{{\it Eureqa} can be purchased at 
\url{https://www.nutonian.com/products/eureqa}.},
implementing an improved version of the generic search algorithm outlined in \cite{schmidt2009distilling}.

We compared the {\it AI Feynman} and {\it Eureqa} algorithms by applying them both to the Feynman Database for Symbolic Regression, allowing a maximum of 2 hours of CPU time per mystery \footnote{The AI Feynman algorithm was run using the hyperparameter settings in Table~\ref{hyperparameterTable}. For {\it Eureqa}, each mystery was run on 4 CPUs. The symbols used in trying to solve the equations were: $+$, $-$, $*$, $/$, constant, integer constant, input variable, 
sqrt, $\exp$, $\log$, $\sin$, $\cos$. To help {\it Eureqa} gain speed, we included the additional functions $\arcsin$ and $\arccos$ only for those mysteries requiring them, and we used only 300 data points (since it does not use a neural network, adding additional data does not help significantly). The time taken to solve an equation using our algorithm, as presented in Tables \ref{EqTableA} and \ref{EqTableB}, corresponds to the time needed for an equation to be solved using a set of symbols that can actually solve it (see Table \ref{opTable}). Equations 1.15.3t and 1.48.2 were solved using the second set of symbols, so the overall time needed for these two equations is one hour larger than the one listed in the tables. Equations I.15.3x and II.35.21 were solved using the 3rd set of symbols, so the overall time taken is two hours larger than the one listed here.
}.
Tables~\ref{EqTableA} and~\ref{EqTableB} show that {\it Eureqa} solved 71\% of the 100 basic mysteries, while {\it AI Feynman} solved 100\%.
Closer inspection of these tables reveal that the greatest improvement of our algorithm over {\it Eureqa} is for the most complicated mysteries, where our neural network enables eliminating variables by discovering symmetries and separability.

\bigskip

The neural network becomes even more important when we rerun {\it AI Feynman} without the dimensional analysis module: it now solves 93\% of the mysteries, and makes very heavy use of the neural network to discover separability and translational symmetries. Without dimensional analysis, many of the mysteries retain variables that appear only raised to some power or in a multiplicative prefactor, and {\it AI Feynman} tends to recursively discover them and factor them out one by one. 
For example, the neural network strategy is used six times when solving
$$F = \frac{Gm_1m_2}{(x_2-x_1)^2+(y_2-y_1)^2+(z_2-z_1)^2}$$
without dimensional analysis: three times 
to discover translational symmetry that replaces $x_2-x_1$, $y_2-y_1$ and $z_2-z_1$ by new variables, once to group together $G$ and $m_1$ into a new variable $a$, once to group together $a$ and $m_2$ into a new variable $b$, and one last time to discover separability and factor out $b$. This shows that although dimensional analysis often provides significant time savings, it is usually not necessary for successfully solving the problem.

Inspection of how {\it AI Feynman} and {\it Eureqa} make progress over time reveals interesting differences.
The progress of {\it AI Feynman} over time corresponds to repeatedly reducing the number of independent variables, and every time this occurs, it is virtually guaranteed to be a step in the right direction. 
In contrast, genetic algorithms such as {\it Eureqa} make progress over time by finding successively better approximations, but there is no guarantee that more accurate symbolic expressions are closer to the truth when viewed as strings of symbols. 
Specifically, by virtue of being a genetic algorithm, {\it Eureqa} has the advantage of not searching the  space of symbolic expressions blindly like our brute force module, but rather with the possibility of a net drift toward more accurate (``fit") equations. The flip side of this is that if {\it Eureqa} finds a fairly accurate yet incorrect formula with a quite different functional form, it risks getting stuck near that local optimum.
This reflects a fundamental challenge for genetic approaches symbolic regression: if the final formula is composed of separate parts that are not summed but combined in some more complicated way (as a ratio, say), then each of the parts may be useless fits on their own and unable to evolutionarily compete. 

\subsection{Dependence on data size}

To investigate the effect of changing the size of the data set, we repeatedly reduced the size of each data set by a factor of 10 until our {\it AI Feynman} algorithm failed to solve it. 
As seen in Tables~\ref{EqTableA} and \ref{EqTableB}, most equations are discovered by the polynomial fit and brute force methods using only 10 data points. 100 data points are needed in some cases, because the algorithm may otherwise overfit when the true equation is complex, ``discovering" an incorrect equation that is too simple.

 As expected, equations that require the use of a neural network to be solved need significantly more data points (between $10^2$ and $10^6$) for the network to be able to learn the mystery function 
accurately enough (i.e. obtaining 
{\rms} accuracy better than $10^{-3}$).
Note that expressions requiring the neural network are typically more complex, so one might intuitively expect them to require 
larger data sets for the correct equation to be discovered without overfitting, even when using alternate approaches such as genetic algorithms. 

\subsection{Dependence on noise level} 
\label{NoiseSec}

Since real data is almost always afflicted with measurement errors or other forms of noise, we investigated the robustness of our algorithm. For each mystery, we added
independent Gaussian random noise to its dependent variable $y$, of standard deviation
$\epsilon\>\>y_{\rm rms}$,
where $y_{\rm rms}$ denotes the {\rms} $y$-value for the mystery before noise has been added. 
We initially set the relative noise level 
$\epsilon=10^{-6}$, then repeatedly multiplied $\epsilon$ by 10 until the {\it AI Feynman} algorithm could no longer solve the mystery.
As seen in Tables~\ref{EqTableA} and~\ref{EqTableB}, most of the equations can still be recovered exactly with an $\epsilon$-value of $10^{-4}$ or less, while almost half of them are still solved for $\epsilon=10^{-2}$. 

For these noise experiments, we adjusted 
the threshold for the brute force and polynomial fit algorithms when the noise level changed, such that not finding a solution at all was preferred over finding an approximate solution. These thresholds were not optimized for each mystery individually, so a better choice of these thresholds might allow the exact equation to be recovered with an even higher noise level for certain equations.
In future work, it will be also be interesting to quantify performance of the algorithm on data with noise added to the independent variables, as well as directly on real-world data. 

\subsection{Bonus mysteries}

The 100 basic mysteries discussed above should be viewed as a training set for our {\it AI Feynman} algorithm, since we made improvements to its implementation and hyper-parameters to optimize performance.
In contrast, we can view the 20 bonus mysteries as a test set, since we deliberately selected and analyzed them only after the {\it AI Feynman} algorithm and its hyper-parameter settings (Table~\ref{hyperparameterTable}) had been finalized.
The bonus mysteries are interesting also by virtue of being significantly more complex and difficult, in order to better identify the limitations our our method.

Table~\ref{EqTableC} sbows that {\it Eureqa} solved only 
15\% of the bonus mysteries, while {\it AI Feynman} solved 90\%. The fact that the success percentage differs more between the two methods for the bonus mysteries than for the basic mysteries reflects the increased equation complexity, which requires our neural network based strategies for a larger fraction of the cases. 

To shed light on the limitations of the {\it AI Feynman} algorithm, it is interesting to consider the two mysteries for which it failed. 
The radiated gravitational wave power mystery was reduced to the form 
$y=-\frac{32 a^2(1+a)}{5b^5}$ by dimensional analysis, corresponding to the string ``$aaa>**bbbbb****/$" in reverse Polish notation (ignoring the multiplicative prefactor $-\frac{32}{5}$). This would require about 2 years for the brute force method, exceeding our allotted time limit. 
The Jackson 2.11 mystery was reduced
to the form 
$a-\frac{1}{4\pi}\frac{a}{b(1-a^2)^2}$
by dimensional analysis, 
corresponding to the string ``$aP0>>>>*\backslash abaa*<aa*<**/*-$" in reverse Polish notation, which would require about 100 times the age of our universe for the brute force method.

It is likely that both of these mysteries can be solved with relatively minor improvements of the our algorithm. 
The first mystery would have been solved had the algorithm not failed to discover that $a^2(1+a)/b^5$ is separable. The large dynamic range induced by the fifth power in the denominator caused the neural network to miss the separability tolerance threshold; potential solutions include temporarily limiting the parameter range or analyzing the logarithm of the absolute value (to discover additive separability). 

If we had used different units in the second mystery, where  $1/4\pi\epsilon$ was replaced by the Coulomb constant $k$, the costly 
$4\pi$-factor (requiring 7 symbols 
``$PPPP+++$" or ``$P0>>>>*$")
would have disappeared.
Moreover, if we had used a different 
set of function symbols that included 
``$Q$" for squaring, then
brute force could quickly have discovered
that $a-\frac{a}{b(1-a^2)^2}$ is solved by ``$aabaQ<Q*/-$".
Similarly, introducing a symbol $\wedge$ denoting exponentiation, enabling the string for $a^b$ to be shortened from 
$``aLb*E"$ to $``ab\wedge"$, would enable brute force to solve many mysteries faster, including Jackson 2.11.

Finally, a powerful strategy that could ameliorate both of these failures would be to add symbols corresponding to parameters that are numerically optimized over. This strategy is currently implemented in {\it Eureqa} but not {\it AI Feynman}, and could make a useful upgrade as long as it is done in a way that does not unduly slow down the symbolic brute force search.
In summary, the two failures of the {\it AI Feynman} algorithm signal not unsurmountable obstacles, but motivation for further work.

In addition, we tested the performance of our algorithm on the mystery functions presented in \cite{mcdermott2012genetic}\footnote{We want to thank the anonymous reviewer who brought this data set to our attention.}. Some  equations appear twice; we included them only once. Our algorithm again outperformed Eureqa, discovering $66.7\%$ of the equations while Eureqa discovered $48.9\%$.
The fact that the AI Feynman algorithm performs less well on this test set than on genuine physics formulas traces back to the fact that most of the equations presented in \cite{mcdermott2012genetic} are rather arbitrary compositions of elementary functions unlikely to occur in real-world problems, thus lacking the symmetries, separability, {\etc} that the neural network part of our algorithm is able to exploit.

\section{Conclusions}
\label{ConclusionsSec}

We have presented a novel physics-inspired algorithm for solving multidimensional analytic regression problems: finding a symbolic expression that matches data from an unknown algebraic function. Our key innovation lies in combining traditional fitting techniques with a
neural-network-based approach that can repeatedly reduce a problem to simpler ones, eliminating dependent variables by discovering properties such as symmetries and separability in the unknown function.

To facilitate quantitative benchmarking of our and other symbolic regression algorithms, we created a freely downloadable database with 100 regression mysteries drawn from the Feynman Lectures on Physics and a bonus set of an additional 20 mysteries selected for difficulty and fame.

\subsection{Key findings}

The pre-existing state-of-the-art symbolic regression software {\it Eureqa} \cite{dubvcakova2011eureqa} discovered 68\% of the Feynman equations and 15\% of the bonus equations, while our {\it AI Feynman} algorithm discovered 100\% and 90\%, respectively, including Kepler's ellipse equation mentioned in the introduction (3rd entry in Table~\ref{EqTableC}).
Most of the 100 Feynman equations could be solved even if the data size was reduced to merely $10^2$ data points or had percent-level noise added, but the most complex equations needing neural network fitting required more data and less noise.

Compared with the genetic algorithm of {\it Eureqa}, the most interesting improvements are seen for the most difficult mysteries where the neural network strategy is repeatedly deployed. 
Here the progress of {\it AI Feynman} over time corresponds to repeatedly reducing the
problem to simpler ones with fewer variables, while {\it Eureqa} and other genetic algorithms are forced to solve the full problem by exploring a vast search space, risking getting stuck in local optima.


\subsection{Opportunities for further work}

Both the successes and failures of our algorithm motivate further work to make it better, and we will now briefly comment on promising improvement strategies.

Although we mostly used the same elementary function options 
(Table~\ref{opTable})
and hyperparameter settings (Table~\ref{hyperparameterTable}) for all mysteries, these could be strategically chosen based on an automated pre-analysis of each mystery. For example, observed oscillatory behaviour
could suggest including $\sin$ and $\cos$ and lack thereof could suggest saving time by excluding them.

Our code could also be straightforwardly integrated  into a larger program discovering equations involving derivatives and integrals, which frequently occur in physics equations. For example, if we suspect that our formula contains a partial differential equation, then the user can simply estimate various derivatives from the data (or its interpolation, using a neural network) and include them in the AI Feynman algorithm as independent variables, thus discovering the differential equation in question.

We saw how, even if the mystery data has very low noise, significant {\it de facto} noise was introduced by imperfect neural network fitting, complicating subsequent solution steps. It will therefore be valuable to explore better neural network  architectures, ideally reducing fitting noise to the $10^{-6}$ level.
This may be easier than in many other contexts, since we do not care if the neural network generalizes poorly outside the domain where we have data: as long as it is highly accurate within this domain, it serves our purpose of correctly factoring separable functions, {\etc}.

 Our brute-force method can be better integrated with a neural network search for hidden simplicity.
 Our implemented symmetry search simply tests if two input variables $a$ and $b$ can be replaced by a bivariate function of them, specifically $+$, $-$, $*$ or $/$, corresponding to length-3 strings
 ``$ab+$",  ``$ab-$",  ``$ab*$" and  ``$ab/$" in Reverse Polish Notation. This can be readily generalized to longer strings involving 2 or more variables, for example bivariate functions
$ab^2$ or $e^a\cos b$.
 
A second example of improved brute-force use is if the neural network reveals that the
function can be exactly solved after setting some variable $a$ equal to something else (say zero, one or another variable). 
A brute force search can now be performed 
in the vicinity of the discovered exact expression: for example, if the expression is valid for $a=0$, the brute force search can insert additive terms that vanish for $a=0$ and multiplicative terms that equal unity for $a=0$, thus being likely to discover the full formula much faster than an unrestricted brute force search from scratch. 

Last but not least, it is likely that marrying the best features from both our method and genetic algorithms can spawn a method that outperforms both. 
Genetic algorithms such as {\it Eureqa} perform quite well even in presense of significant noise, whether they output not merely one hopefully correct formula, but rather a Pareto frontier, a sequence of increasingly complex formulas that provide progressively better accuracy. Although it may not be clear which of these formulas is correct, it is more likely that the correct formula is {\it one of them} than any particular one that an algorithm might guess. When our neural network identifies separability, a so generate Pareto frontier could thus be used to generate candidate formulas for one factor, after which each one could be substituted back and tested as above, and the best solution to the full expression would be retained. Our brute force algorith can similarly be upgraded to return a Pareto frontier instead of a single formula.

In summary, symbolic regression algorithms are getting better, and are likely to continue improving.
We look forward to the day when, for the first time in the history of physics, a computer, just like Kepler, discovers a useful and hitherto unknown physics formula through symbolic regression!

\bigskip

{\bf Acknowledgements:} 
We thank Rustin Domingos, Zhiyu Dong, Michael Skuhersky, Andrew Tan and Tailin Wu for helpful comments,  and the Center for Brains, Minds, and Machines (CBMM) for hospitality.
{\bf Funding:} This work was supported by The Casey and Family Foundation, the Ethics and Governance of AI Fund, the Foundational Questions Institute, the Rothberg Family Fund for Cognitive Science and the Templeton World Charity Foundation, Inc. The opinions expressed in this publication are those of the authors and do not necessarily reflect the views of the Templeton World Charity Foundation, Inc.
{\bf Author contributions}: Concept, supervision, project management: M.T.
Design of methodology, programming, experimental experimental validation, data curation, data analysis, validation, manuscript writing: S.U. and M.T.
{\bf Competing interests:} The authors declare that they have no competing interests. 
{\bf Data and materials availability:} All data needed to evaluate the conclusions in the paper are present in the paper, 
at \url{https://space.mit.edu/home/tegmark/aifeynman.html}
and at \url{https://github.com/SJ001/AI-Feynman}. Any additional datasets, analysis details, and material recipes are available upon request.

\bigskip
\bigskip

\begin{algorithm}[h]
\caption{\textbf{AI Feynman: Translational Symmetry}}
\label{tran_sym}
\begin{algorithmic}
\STATE  {\bfseries Require} Dataset $\D=\{(\x, \y)\}$
\STATE  {\bfseries Require} \textbf{net}: trained neural network
\STATE {\bfseries Require $\mathbf{NN_{error}}$}: the neural network validation error

\STATE a = 1 
\STATE \textbf{for} i \textbf{in} len(\textbf{x}) \textbf{do}:
\STATE \ \ \ \ \textbf{for} j \textbf{in} len(\textbf{x}) \textbf{do}:
\STATE \ \ \ \ \ \ \ \ \textbf{if} i $<$ j:
\STATE \ \ \ \ \ \ \ \ \ \ \ \ $x_t$ = \textbf{x}
\STATE \ \ \ \ \ \ \ \ \ \ \ \ $x_t$[i] = $x_t$[i] + a
\STATE \ \ \ \ \ \ \ \ \ \ \ \ $x_t$[j] = $x_t$[j] + a
\STATE \ \ \ \ \ \ \ \ \ \ \ \ \textbf{error} = RMSE(\textbf{net}(\textbf{x}),\textbf{net}($x_t$)) 
\STATE \ \ \ \ \ \ \ \ \ \ \ \ \textbf{error} = \textbf{error}/RMSE(\textbf{net}(\textbf{x}))
\STATE \ \ \ \ \ \ \ \ \ \ \ \ if \textbf{error} $<$ $7 \times \mathbf{NN_{error}}$:
\STATE \ \ \ \ \ \ \ \ \ \ \ \ \ \ \ \ $x_t[i]$ = $x_t[i]-x_t[j]$ 
\STATE \ \ \ \ \ \ \ \ \ \ \ \ \ \ \ \ $x_t$ = delete($x_t$,j) 
\STATE \ \ \ \ \ \ \ \ \ \ \ \ \ \ \ \ \textbf{return} $x_t$, i, j
\end{algorithmic}
\end{algorithm}

\begin{algorithm}[h]
\caption{\textbf{AI Feynman: Additive Separability}}
\label{alg:additive_separability}
\begin{algorithmic}
\STATE  {\bfseries Require} Dataset $\D=\{(\x, \y)\}$
\STATE  {\bfseries Require} \textbf{net}: trained neural network
\STATE {\bfseries Require $\mathbf{NN_{error}}$}: the neural network validation error

\STATE $x_{eq}$ = \textbf{x} 
\STATE \textbf{for} i \textbf{in} len(\textbf{x}) \textbf{do}:
\STATE \ \ \ \  $x_{eq}[i]$ = mean(\textbf{x}$[i]$)

\STATE \textbf{for} i \textbf{in} len(\textbf{x}) \textbf{do}:

\STATE \ \ \ \ \textbf{c} = combinations([1,2,...,len(\textbf{x})],i)
\STATE \ \ \ \ \textbf{for} $\mathrm{idx_1}$ \textbf{in} c \textbf{do}:
\STATE \ \ \ \ \ \ \ \ $x_1$ = \textbf{x} 

\STATE \ \ \ \ \ \ \ \ $x_2$ = \textbf{x} 

\STATE \ \ \ \ \ \ \ \ $\mathrm{idx_2}$ = k \textbf{in} [1,len(\textbf{x})] \textbf{not in} $\mathrm{idx_1}$
\STATE \ \ \ \ \ \ \ \ \textbf{for} j \textbf{in} $\mathrm{idx_1}$:
\STATE \ \ \ \ \ \ \ \ \ \ \ \ $x_1[j]$ = mean(\textbf{x}$[j]$)
\STATE \ \ \ \ \ \ \ \ \textbf{for} j \textbf{in} $\mathrm{idx_2}$:
\STATE \ \ \ \ \ \ \ \ \ \ \ \ $x_2[j]$ = mean(\textbf{x}$[j]$)
\STATE \ \ \ \ \ \ \ \ \textbf{error} = RMSE(\textbf{net}(\textbf{x}),\textbf{net}($x_1$)+\textbf{net}($x_2$)-\textbf{net}($x_{eq}$)) 
\STATE \ \ \ \ \ \ \ \ \textbf{error} = \textbf{error}/RMSE(\textbf{net}(\textbf{x}))
\STATE \ \ \ \ \ \ \ \ \textbf{if} \textbf{error} $ <$ $10 \times \mathbf{NN_{error}}$:
\STATE \ \ \ \ \ \ \ \ \ \ \ \ $x_1$ = delete($x_1$,$\mathrm{index_2}$)
\STATE \ \ \ \ \ \ \ \ \ \ \ \ $x_2$ = delete($x_2$,$\mathrm{index_1}$)
\STATE \ \ \ \ \ \ \ \ \ \ \ \ \textbf{return} $x_1$, $x_2$, $\mathrm{index_1}$, $\mathrm{index_2}$
\end{algorithmic}
\end{algorithm}

\bibliography{feynman}

\end{document}